\documentclass[12pt]{article}

\def\lsim{\mathrel{\rlap {\raise.5ex\hbox{$ < $}}
{\lower.5ex\hbox{$\sim$}}}}
\def\gsim{\mathrel{\rlap {\raise.5ex\hbox{$ > $}}
{\lower.5ex\hbox{$\sim$}}}}
\begin{document}

\begin{titlepage}

\begin{centering}
\vspace*{1cm}
\hspace{9cm} hep-th/0205009
\vspace{.6in}

{\bf{Graviton localization and Newton law for a $dS_4$ brane in a
$5D$ bulk} }\\ \vspace{1.5 cm} {A. Kehagias$^{1,2}$
 and K. Tamvakis$^{3}$}\\
\vskip .5cm
{$^1 $\it{Athens National Technical University,
Zografou, GREECE}\\
\vskip .5cm
{$^2$ Institute of Nuclear Physics, N.C.R.P.S. Democritos,\\
 GR-15310,
Athens, GREECE}\\
\vskip .5cm
{$^3$\it {Physics Department, University of Ioannina\\
45110 Ioannina, GREECE}}}\\

\vspace{1cm}
{\bf Abstract}\\
\end{centering}
\vspace{.1in}
We consider an $dS_4$ brane embedded in a five-dimensional bulk
with a positive, vanishing or negative bulk cosmological constant and
derive
the localized graviton spectrum that consists of a
normalizable zero-mode separated by a gap from a continuum of massive
states.
We estimate the massive sector contribution to the static potential at short distances and find that only in
 the case of a negative bulk cosmological constant there is a range,
determined by the effective four-dimensional and the bulk
cosmological constants, where the conventional Newton law is
valid.

\vspace{3.5cm}
\begin{flushleft}

May 2002
\end{flushleft}
\hrule width 6.7cm \vskip.1mm{\small \small}
 \end{titlepage}

 Models of extra
dimensions, in which the standard four dimensions correspond to
 a brane embedded in a higher dimensional bulk have been extensively studied
 in the last few years as a solution to the hierarchy problem\cite{D1}. Soon
 it was realized that the extra dimensions could be large. Non-compact internal spaces have been employed, although with a
limited success, from the early days of the KK programme
\cite{Gell-Mann:1985if},\cite{Nicolai:jg}.
 Their relation to gauged supergravities has
also been pointed out
\cite{Kehagias:1999ju},\cite{Skenderis:1999mm},\cite{DeWolfe:1999cp}.
Independently of how such branes are formed
\cite{Kehagias:2000au},  brane models in non-compact spaces
require the trapping of gravitational degrees of freedom on the
  brane \cite{Randall:1999ee},\cite{RS}. Graviton localization results from the existence of
a normalizable zero mode, although less straightforward situations could arise, as in
 the case of a $AdS_4$ brane embedded in $AdS_5$ bulk where gravity is localized
despite the fact that a
   zero mode is absent\cite{KR}.

Although curved branes embedded in a higher dimensional bulk have been considered before \cite{KR}--\cite{Ito:2002qp},
 the case of a de Sitter brane ($dS_4$) is particularly interesting both for {\textit{phenomenological}}
 as well as theoretical reasons.
In this article we consider the three possible cases of embedding
an $dS_4$ brane in a five dimensional bulk having positive
($\Lambda>0$), vanishing ($\Lambda=0$) or negative ($\Lambda<0$)
bulk cosmological constant. In all these cases the graviton
spectrum consists of a zero-mode separated by a gap
 from a continuum of massive modes. Although the contribution of
the zero-mode to the static gravitational potential is always
 Newtonian,
the massive modes introduce a five-dimensional behaviour which
 dominates
in all cases except the case of negative
bulk cosmological constant ($AdS_5$) in which ordinary gravity is
 possible
and five-dimensional behaviour shows up only at very short distances.

   We start from the five-dimensional Einstein action with a
cosmological
constant
$\Lambda$ and a brane of tension $\sigma$
\begin{equation}{\cal{S}}=\int d^5x\sqrt{-g}\left\{2M^3R-\Lambda\right\}
-\sigma\int d^4x\sqrt{-\overline{g}}
\end{equation}
 and introduce a general metric ansatz
\begin{equation}ds^2=e^{2A(y)}
\overline{g}_{\mu\nu}(x)dx^{\mu}dx^{\nu}+dy^2
\end{equation}
with a four-dimensional metric $\overline{g}_{\mu\nu}(x)$
fixed to correspond to a de Sitter space
   \begin{equation}\overline{g}_{\mu\nu}=\left(\begin{array}{cc}
   -1\,&\,0\\
   0\,&\,e^{2Ht}\delta_{ij}
   \end{array}\right)\end{equation}
   The resulting equations of motion are
   \begin{equation}-3H^2e^{-2A}+3\left(A^{''}+(A')^2\right)=
   -\frac{\Lambda}{4M^3}-\frac{\sigma}{4M^3}
\delta(y)\end{equation}
   \begin{equation}-6H^2e^{-2A}+6(A')^2 =-\frac{\Lambda}{4M^3}\end{equation}
   These equations can easily be solved
   \cite{DeWolfe:1999cp},\cite{Cvetic:1993xe},\cite{Kaloper:1999sm}, for
either sign of $\Lambda$.

{\textbf{ \noindent\\
Positive bulk cosmological constant $\bf{\Lambda}>0$.}}
In
the case $\Lambda>0$, we have the
   warp factor\footnote{An equivalent representation of the metric as
$ds^2=e^{2A(z)}\left(\overline{g}_{\mu\nu}dx^{\mu}dx^{\nu}+dz^2\right)$
   in terms of $z=\int_0^{y}dye^{-A(y)}$ gives the warp factor expression
   $$e^{2A(z)}=e^{-2H|z|}\frac{(1+\tan^2(ny_0/2))}{(1+\tan^2(ny_0/2)
e^{-2H |z|})^2}$$}
   \begin{equation}e^{A(y)}=\frac{\sin\left(n(y_0-|y|)\right)}{\sin(ny_0)}
\end{equation}
   where, we have introduced
   \begin{equation}n^2\equiv
   \frac{\Lambda}{24M^3}\,\,\,,\,\,\,y_0\equiv
\frac{1}{n}\cot^{-1}(\frac{\sigma}{24M^3n})\end{equation}
   The metric ansatz parameter $H$ is related to the action parameters
through the relation
   \begin{equation}H^2=n^2+\left(\frac{\sigma}{24M^3}\right)^2=
\frac{n^2}{\sin^2(ny_0)}\,\,\,\Rightarrow (H^2>n^2)\end{equation}
   Either expression implies that $H^2$ receives a positive
   contribution
from the bulk cosmological constant.
   The Ricci scalar in this case is
   \begin{equation}R=\frac{2\sigma}{M^3}\delta(y)+20 n^2\end{equation}
   corresponding to a de Sitter bulk.

   The graviton spectrum can be obtained by performing the variation
$\delta g_{MN}=\delta_M^{\mu}\delta_N^{\nu}h_{\mu\nu}(x,y)$ and
   introducing the ansatz\footnote{This can be done in the harmonic
gauge.
The tracelessness-transversality
   conditions are
$\eta_{\mu}^{\mu}=\overline{\nabla}_{\mu}\eta_{\nu}^{\mu}=0$.}
$h_{\mu\nu}=\psi(y)\eta_{\mu\nu}(x)$. The resulting graviton equations
   are
   \begin{equation}-\frac{1}{2}\psi^{''}(y)+\left(A^{''}+2(A')^2\right)
\psi(y)=\frac{m^2}{2}e^{-2A}\psi(y)\end{equation}
   \begin{equation}-\frac{1}{2}\overline{\nabla}^2\eta_{\mu\nu}(x)+H^2
\eta_{\mu\nu}(x)=-\frac{m^2}{2}\eta_{\mu\nu}(x)\end{equation}
   There is a normalizable\footnote{Normalizability corresponds to
$\int dy e^{-2A}|\psi(y)|^2<\infty$.} zero mode ($m^2=0$)
   \begin{equation}\psi_0(y)=e^{2A(y)}=\frac{\sin^2\left(n(y_0-|y|)\right)}{\sin^2(ny_0)}
\end{equation}
whereas the massive spectrum can be determined from
\begin{equation} -\frac{1}{2}\psi^{''}(y)+
n^2\left\{-2+\frac{(1-m^2/2H^2)}{\sin^2\left(n(y_0-|y|)\right)}
\right\}\psi(y)=\frac{\sigma}{12M^3}\psi(0)\delta(y)\end{equation}
Acceptable solutions exist for $m^2>9H^2/4$. They are, up to a
dimensionless multiplicative constant,
\begin{equation} \frac{\sqrt{n}}{\sin^2\left(n(y_0-|y|)\right)}
{\,}_2F_1\left(5/4\mp 3\delta/4,\,5/4\pm 3\delta/4;\,3;\,\sin^{-2}
\left(n(|y|-y_0)\right)\right)\end{equation}
where $\delta^2=4m^2/9H^2-1$.

{\textbf{\noindent\\
 Vanishing bulk cosmological constant $\bf{\Lambda}=0$.} }
The case of vanishing bulk cosmological constant $\Lambda=0$ is
essentialy a case studied long ago\cite{V} and corresponds to
the warp factor\footnote{The metric takes the form $e^{-2H|z|}
\left\{\overline{g}_{\mu\nu}(x)dx^{\mu}dx^{\nu}+dz^2\right\}$,
with
$\overline{g}_{\mu\nu}=Diag (-1,\,e^{2Ht}\delta_{ij})$, in terms
of the variable $|z|=-\frac{1}{H}\ln (1-H|y|)$.}
\begin{equation} e^{A(y)}=1-H|y|\end{equation}
and the relation
\begin{equation} H=\frac{\sigma}{24M^3}\end{equation}
The four-dimensional effective Planck mass is in this case
$M_P^2=2M^3/3H$.
The graviton localization equation is
\begin{equation}-\frac{1}{2}\psi^{''}(y)+H^2\frac{(1-m^2/2H^2)}{(1-H|y|)^2}
\psi(y)=2H\psi(0)\delta(y)\end{equation}
There is a normalizable zero-mode
\begin{equation}\psi_0(y)=\sqrt{\frac{3H}{2}}(1-H|y|)^2\end{equation}
and a continuum of massive delta-function normalizable states for $4m^2>9H^2$
\begin{equation}\psi_{\delta}(y)=\sqrt{\frac{3H}{8\pi}}(1-H|y|)^{1/2}
\left\{(1-H|y|)^{3i\delta /2}+\left(\frac{i\delta-1}{i\delta +1}\right)
(1-H |y|)^{-3i\delta /2}\right\}\end{equation}
where $\delta$ is the previously defined parameter. This case offers
the advantage of having elementary analytic expressions for the
graviton
{\textit{wave functions}} in
contrast to the previous case where the massive continuum states were
Hypergeometric functions.

{\textbf{ \noindent\\
Negative bulk cosmological constant ${\bf{\Lambda}}<0$.}}
In the case of negative bulk cosmological constant, the warp
factor comes out to be\footnote{In terms of a $z$-variable, defined
as in the $\Lambda>0$ case, the warp factor takes the form
$$e^{2A(z)}=\frac{\left(1-\tanh^2(\nu y_0/2)\right)}{\left(1-
\tanh^2(\nu y_0/2)e^{-2H|z|}\right)}e^{-2H|z|}$$}
\begin{equation}e^{A(y)}=
\frac{\sinh\left(\nu(y_0-|y|)\right)}{\sinh(\nu y_0)}\end{equation}
where, we have defined
\begin{equation}\nu^2\equiv -\frac{\Lambda}{24M^3}\,\,,\,\,\,\,
\,\,\nu y_0=\coth^{-1}(\sigma/24M^3\nu)\end{equation}
The relation between action and metric-ansatz parameters is
\begin{equation}H^2=\left(\frac{\sigma}{24M^3}\right)^2-\nu^2=
\frac{\nu^2}{\sinh^2(\nu y_0)}\,\,\,\Rightarrow (H^2<\nu^2)\end{equation}
In this case the bulk is asymptotically $AdS$ with a Ricci scalar
\begin{equation}R=\frac{2\sigma}{3M^3}\delta(y)-20\nu^2\end{equation}

The spectrum consists of a normalizable zero mode
$\psi_0(y)=e^{2A(y)}$
and a continuum of massive states with masses $4m^2>9H^2$ and
wave-functions corresponding to the solutions of
\begin{equation}-\frac{1}{2}\psi^{''}(y)+\nu^2\left\{2+\frac{(1-m^2/2H^2)}{
\sinh^2\left(\nu(y_0-|y|)\right)}\right\}\psi(y)=\frac{\sigma}{12M^3}\psi(0)
\delta(y)\end{equation}
These solutions are, up to a dimensionless multiplicative constant,
\begin{equation}
\psi(y)\sim
\frac{\sqrt{\nu}}{\sinh^2\left(\nu(y_0-|y|)\right)}{\,}_2
F_1\left(5/4\mp 3i\delta/4,\,5/4\pm
3i\delta/4;\,3;\,-\sinh^{-2}\left(\nu(y_0-|y|)
\right)\right)\end{equation}

{\textbf{ \noindent\\ The gravitational potential.}} A most
relevant question for all the above cases is the question of the
{\textit{gravitational potential}} created by a unit mass on the
brane. This potential is directly related to the $00$ component of
the graviton, namely $V(x,y)=\frac{1}{2}h_{00}(x,y)$. (Newton law
in de-Sitter space \cite{Tsamis:1992xa} and time-depended
backgrounds have been discussed in \cite{Iliopoulos:1998wq}.) For
a point particle of mass $\mu$ at a point $\vec{x}=0$ on the brane
it satisfies the equation
\begin{equation}-\frac{1}{2}e^{-2A}\overline{\nabla}^2V+H^2V-\frac{1}{2}
V^{''}+\left(A^{''}+2(A')^2\right)V=-\lambda \delta^{(3)}(x)\delta(y)\end{equation}
where $\lambda=\mu/M^3$. Expanding the potential in terms of the eigenfunctions $\psi_0(y)$ and
$\psi_{\delta}(y)$ as
\begin{equation}V(x,y)=\phi_0(x)\psi_0(y)+\int_0^{\infty}d\delta\,
\phi_{\delta}(x)\psi_{\delta}(y)\end{equation}
Substituting into the equation for $V$, we obtain for the coefficient
functions $\phi_{\delta}(x)$ and $\phi_0(x)$
\begin{eqnarray}
&&-\frac{1}{2}\overline{\nabla}^2\phi_{\delta}(x)+
(H^2+\frac{m^2}{2})\phi_{\delta}(x)=-\lambda
\psi_{\delta}^{*}(0)\delta^{(3)}(x)\, , \nonumber \\
&&-\frac{1}{2}
\overline{\nabla}^2\phi_0(x)+H^2\phi_0(x)=-\lambda
\psi_0(0)\delta^{(3)}(x)
\end{eqnarray}
The solutions to these equations can be written in terms of the de
Sitter propagator ${\cal{S}}(x,x')$ defined as
\begin{equation}{\cal{S}}^{-1}(x,x')\equiv
\left\{\left(-\frac{\partial^2}
{\partial\tau^2}+\vec{\nabla}^2\right)-2\frac{H}{(1-H\tau)}\frac{\partial}
{\partial \tau}-\frac{m^2}{(1-H\tau)^2}\right\}\delta^{(4)}(x-x')
\end{equation}
where we have introduced the conformal time
$$\tau=\frac{1}{H}(1-e^{-Ht})\,\,\,\,,\,\,\,\,-\infty<\tau<H^{-1}$$
as in \cite{Higuchi:gz}, \cite{Higuchi:2001uv}.
The coefficient functions are
\begin{equation}\phi_0(x)=2\lambda \psi_0(0)\int
\frac{d\tau'}{(1-H\tau')^2}
{\cal{S}}_0(x,0,\tau')\,\,\,,\,\,\,\,\phi_{\delta}(x)=2\lambda
\psi_{\delta}^{*}(0)\int \frac{d\tau'}{(1-H\tau')^2}{\cal{S}}_{\delta}
(x,0,\tau')\end{equation}
The expression of the gravitational potential on the brane is
$$V(x,0)=2\lambda\psi_0^2(0)\int_{-\infty}^{H^{-1}}\frac{d\tau'}{(1-H\tau')^2}
{\cal{S}}_0(\tau,\vec{x};\tau',0)
+$$
\begin{equation}2\lambda\int_0^{\infty}d\delta|\psi_{\delta}(0)|^2
\int_{-\infty}^{H^{-1}}\frac{d\tau'}{(1-H\tau')^2}
{\cal{S}}_{\delta}(\tau,\vec{x};\tau',0)
\end{equation}

The
differential equation satisfied by the propagator can be transformed
into a differential
equation with respect to the geodesic distance, namely
$$z(1-z){\cal{S}}^{''}(z)+2(1-2z){\cal{S}}'(z)-\frac{m^2}{H^2}{\cal{S}}(z)
=0$$
where the  geodesic distance $z$ is defined, with  $\gamma\equiv
H^{-1}-\tau$,
as\begin{equation}z\equiv \frac{(\gamma+\gamma')^2-
(\vec{x}-\vec{x}^{\,\prime})^2}{4\gamma\gamma'}.
\end{equation}
Solving the above equation, we find that the
 propagator \cite{A} is
\begin{equation}{\cal{S}}_{\delta}(z)=\frac{H^2}{16\pi^2}|
\Gamma(3/2+3i\delta /2)|^2{\,}_2F_1(3/2+3i\delta/2,\,3/2-3i\delta/2;\,2;\,z)\end{equation}
The massless propagator can be directly obtained in an elementary form
from the differential equation it satisfies, namely
\begin{equation}{\cal{S}}_0(z)=\frac{iH^2}{16\pi^2}\left(\frac{1}{1-z-i
\epsilon}-\frac{1}{z+i\epsilon}+2\ln
\left(\frac{z+i\epsilon}{1-z-i\epsilon}
\right)\right)\end{equation}
The contribution of the zero mode to the potential can be integrated\footnote{The replacement of zero with $\gamma$ as a lower integration limit is kinematical.}
and gives,
in the $R\rightarrow 0$ limit a real part that is just
Newtonian
\begin{equation}V_0(x)=\frac{2\lambda}{H^2}\psi_0^2(0)\int_{\gamma}^{\infty}
\frac{d\gamma'}{{\gamma'}^2}{\cal{S}}_0(z)\sim -\frac{\lambda
\psi_0^2(0)}{4\pi }\frac{1}{R}.
\end{equation}
 In the $\Lambda=0$ case, setting
$\lambda=\mu/M^3$, we get
$$V_0(R)\sim -\frac{3\mu H}{8\pi M^3}\frac{1}{R}=-\frac{\mu}{4\pi
M_P^2}\frac{1}{R}$$
as we expect.

The contribution of the massive continuum is expressed as a double integral
\begin{equation}V_m(x)=\frac{2\lambda}{H^2}\int_0^{\infty}d\delta
|\psi_{\delta}(0)|^2\int_{\gamma}^{\infty}\frac{d\gamma'}{{\gamma'}^2}
{\cal{S}}_{\delta}(z)\end{equation}
In the short-distance limit $HR<<1$, the propagator is expected to behave as a flat propagator. Thus, for $R<<H^{-1}$, but still $m^{-1}<R$, we can roughly approximate the
time
integral with its massless value suppressed by an exponential, namely
\begin{equation}\int_{\gamma}^{\infty}\frac{d\gamma'}{{\gamma'}^2}
{\cal{S}}_{\delta}(z)\sim -\frac{H^2}{8\pi R}e^{-mR}\end{equation}
Thus,
\begin{equation}V_m(R)\sim -\frac{\lambda}{4\pi
R}\int_0^{\infty}d\delta |\psi_{\delta}(0)|^2e^{-mR}
\end{equation}

In the case $\Lambda=0$, we have $|\psi_{\delta}(0)|^2=\frac{3H}{2\pi}
(\delta^2/(1+\delta^2))$. Setting $\xi=mR>1$, we get $\delta\sim
2\xi/3(HR)
(1+O(HR)^2)$ and $|\psi_{\delta}(0)|^2\sim \frac{3H}{2\pi}(1+O(HR)^2)$.
Thus, the massive states contribution to the potential
can be approximated with
\begin{equation}V_m^{(\Lambda=0)}(R)\sim -\frac{3\lambda H}{8\pi R}
\left(\frac{1}{6\pi}\frac{{\cal{C}}_0}{HR}+O(HR)\right)
\end{equation} where ${\cal{C}}_0=\int
d\xi\,e^{-\xi}(1+O(HR))$. The total potential
 for this case is
\begin{equation}V^{(\Lambda=0)}(R)\sim -\frac{3\lambda H}{8\pi R}
\left(1+\frac{1}{6\pi}\frac{{\cal{C}}_0}{HR}+O(HR)\right)=
-\frac{\mu}{4\pi
M_P^2}\frac{1}{R}\left(1+\frac{{\cal{C}}_0}{6\pi}\frac{1}
{HR}+O(HR)\right) \label{L0}
\end{equation}

In the case $\Lambda>0$, in order to proceed further in the expression
for
$V_m$ we note that, up to a dimensionless constant,
$$|\psi_{\delta}(0)|^2=\frac{n}{\sin^4(ny_0)}\left|{\,}_2F_1\left(5/4\mp
3i\delta/4,
\,5/4\pm 3i\delta/4;\,3;\,\sin^{-2}(n y_0)\right)\right|^2$$
$$=H\left(\frac{H}{n}\right)^3\left|F(\delta,\,H^2/n^2)\right|^2$$
Thus,
\begin{equation}V_m(R)\sim -\frac{\lambda H}{4\pi
R}\left(\frac{H}{n}\right)^3
\int_0^{\infty}d\delta \left|F(\delta, H^2/n^2)\right|^2e^{-mR}\end{equation}
Setting
$$\int d\delta\left|F(\delta, H^2/n^2)\right|^2e^{-mR}\sim
\frac{2}{3(HR)}\int d\xi
\left|F(\xi/(HR), H^2/n^2)\right|^2e^{-\xi}\equiv \frac{3{\cal{C}}_+ (HR, H/n)}{2(HR)}$$
we can write the full potential as
\begin{equation}V^{(\Lambda>0)}(R)\sim  -\frac{3\lambda H}{8\pi R}
\left(1+\left(\frac{H}{n}\right)^3\frac{{\cal{C}}_+}{HR}\right)
 \label{Lbig}
 \end{equation}
Unfortunately, in both of the above cases ($\Lambda\geq 0$) there
can be no argument in favour of the suppression of the
{\textit{``corrections"}} due to the massive states. Taking into
account the increasing behaviour on $H/n$, these corrections
dominate for $R<H^{-1}$ and the five dimensional behaviour is
dominant in the potential which is nowhere Newtonian. It should be
noted that this behaviour is expected since in both the above
cases the effective size of the fifth dimension is $y_0 \sim
H^{-1}$.

This is in sharp contrast to the case of negative bulk cosmological
constant ($\Lambda<0$). In that case, we have, up to
a dimensionless multiplicative constant,
$$|\psi_{\delta}(0)|^2=\frac{\nu}{\sinh^4(\nu
y_0)}|{\,}_2F_1\left(5/4\mp 3i
\delta/4,\,5/4\pm 3i\delta/4;\,3;\,
-\sinh^{-2}(\nu y_0)\right)|^2$$
$$=H\left(\frac{H}{\nu}\right)^3\left|F(\delta,-H^2/\nu^2)\right|^2$$
Thus,
\begin{equation}V_m(R)\sim -\frac{\lambda H}{4\pi
R}\left(\frac{H}{\nu}
\right)^3\int_0^{\infty}d\delta \left|F(\delta, -H^2/\nu^2)\right|^2e^{-mR}\end{equation}
Setting
$$\int d\delta\left|F(\delta, -H^2/\nu^2)\right|^2e^{-mR}\sim \frac{2}
{3(HR)}\int d\xi \left|F(\xi/(HR), -H^2/\nu^2)\right|^2e^{-\xi}\equiv
 \frac{3{\cal{C}}_- (HR, H/\nu)}{2(HR)}$$
we can write the full potential as
\begin{equation}V^{(\Lambda<0)}(R)\sim  -\frac{3\lambda H}{8\pi R}
\left(1+\left(\frac{H}{\nu}\right)^3\frac{{\cal{C}}_-}{HR}\right)
\label{Lles}
\end{equation} Here, since the ratio $H/\nu$ can be
chosen to be as small as desired, taking into account the
decreasing behaviour of ${\cal{C}}_{-}$ as a function of it,
 we can easily argue that there is a range
$$\nu^{-1}<<R<H^{-1} $$ where the Newtonian term due to the zero
mode dominates and gravity is ordinary. Five-dimensional behaviour
comes in at shorther distances $R<<\nu^{-1}$. Note that  this
behaviour goes along with the small size  of the extra dimension
in the case $H<\nu$ which is $y_0 \sim (1/ \nu) \log (\nu / H)$.

In conclusion, we have considered curved $dS_4$ branes in a 5D bulk and in particular,
 graviton localization and  the existence of a Newtonian limit. We looked at the
static gravitational potential at short distances $HR <<1$, given
approximatelly by the equations (\ref{L0}), (\ref{Lbig}) and
(\ref{Lles}) for zero, positive and negative  bulk cosmological
constant, respectively. In all cases, there are
{\textit{corrections}} to the 4D Newton law due
 to the massless graviton as a result of massive gravitational KK
states. Consistency with observations requires the suppression
 of these corrections. However, this is met only the case of negative bulk cosmological constant, where
 we have the conventional Newton law for distances
greater than $\nu^{-1}$ and smaller than $H^{-1}$. In all other cases,
the 5D nature of the background dominates leading to a 5D
gravitational interaction on the brane.

{\textbf{\noindent\\
Acknowledgments}}

One of us (K.T.) acknowledges useful conversations with P. Kanti,
I. Olasagasti, A. Petkou and A. Polychronakos as well as the travelling support of the RTN program with
contract HPRN-CT-2000-00152. This work is
partially supported by the RTN contracts HPRN-CT-2000-00148,
HPRN-CT-2000-00122 and  HPRN-CT-2000-00131.

\end{document}